\documentclass[a4paper,12pt]{article}

\usepackage[T1]{fontenc}
\usepackage{epstopdf}

\usepackage{graphicx}

\usepackage{hyperref}
\usepackage{color}
\renewcommand \UrlFont{\color{blue}\rmfamily}

\begin{document}

\title{
Complex Dynamics of the Implicit Maps
\\Derived from Iteration of Newton and Euler Methods
\thanks{Supported by \href{https://rscf.ru/}{\UrlFont Russian Science Foundation},
Grant No. \href{https://rscf.ru/project/21-12-00121/}{\UrlFont 21-12-00121}.}}

\author{Andrei~A.~Elistratov, Dmitry~V.~Savin$^1$,
Olga~B.~Isaeva$^{1,2}$}

\maketitle\begin{center} \emph{Saratov State University
\\ Astrahanskaya 83 Saratov,
410026, Russia }\end{center}

\date{}
\maketitle\begin{center} \emph{
Kotel'nikov Institute of Radio-Engineering and Electronics of RAS, Saratov Branch \\
Zelenaya 38, Saratov,
410019, Russia}\end{center}

\maketitle              

\begin{abstract}
Special exotic class of dynamical systems~--- the implicit maps~--- is 
considered. Such maps, particularly, can appear as a result of using 
of implicit and semi-implicit iterative numerical methods. In the 
present work we propose the generalization of the well-known Newton-Cayley 
problem. Newtonian Julia set is a fractal boundary on the complex plane,
 which divides areas of convergence to different roots of cubic nonlinear 
complex equation when it is solved with explicit Newton method. We consider 
similar problem for the relaxed, or damped, Newton method, and obtain the 
implicit map, which is non-invertible both time-forward and time-backward. 
It is also possible to obtain the same map in the process of solving of 
certain nonlinear differential equation via semi-implicit Euler method. 
The nontrivial phenomena, appearing in such implicit maps, can be considered, 
however, not only as numerical artifacts, but also independently. From the 
point of view of theoretical nonlinear dynamics they seem to be very 
interesting object for investigation. Earlier it was shown that implicit 
maps can combine properties of dissipative non-invertible and Hamiltonian 
systems. In the present paper strange invariant sets and mixed dynamics of the 
obtained implicit map are analyzed.

keywords: Julia set, Hamiltonian system, implicit map

\end{abstract}

\section*{Introduction}
\begin{figure}
\includegraphics[width=\textwidth]{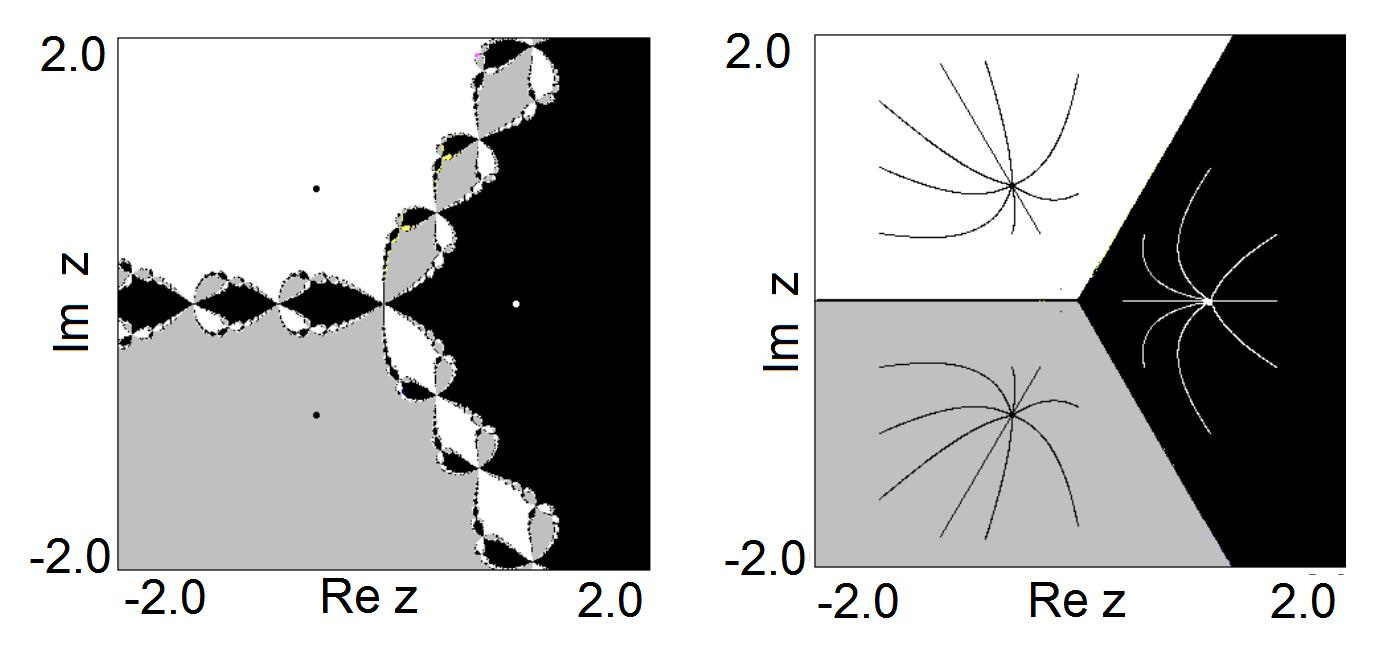}
\caption{
Newtonian Julia set~--- fractal boundary between white, black and gray 
regions, which are the basins of attraction of three roots of the 
equation~(\ref{f3}) with~$c=-1$ (left panel) and Newtonian pie~---  phase plane of the
flow dynamical 
system~(\ref{ode}) (right panel): three roots of~(\ref{f3}) 
are the stable nodes of the ODE~(\ref{ode}), its basins of attraction are 
divided by the smooth separatrix.
Euler discretization~(\ref{eul}) of~(\ref{ode}) results in emergence of fractal basin boundary, 
like one on the left panel, which degenerates to the true smooth linear 
separatrix at~$h\rightarrow 0$.} 
\label{fig0}
\end{figure}

One of the wide-known examples of fractal sets~--- Newtonian Julia set 
(Fig.~\ref{fig0})~---
arises as a boundary of areas of convergence to different roots of the 
cubic polynomial equation on the complex plane
\begin{equation}\label{f3}
\phi(z)=z^3+c=0,
\end{equation}
when it is solved with the Newton method~\cite{Peitgen}. This problem, 
first suggested by Cayley~\cite{Cayley}, allows generalization, if relaxed, 
or damped Newton method
\begin{equation}\label{newt}
z_{n+1}=z_n-h\frac{\phi(z_n)}{\phi'(z_n)}
=z_n-h\frac{z_n^3+c}{2z_n^2}
\end{equation}
is used~\cite{Mclaughlin,Magrenan}. At the same time the iteration 
process~(\ref{newt}) can be considered as a 
trivial discretization of the ordinary differential equation 
\begin{equation}\label{ode}
\dot{z}=f(z)
\end{equation}
with the Euler method
\begin{equation}\label{eul}
z_{n+1}=z_n+h f(z_n),
\end{equation}
where
\begin{equation}\label{fun}
f(z)=-\frac{\phi(z)}{\phi'(z)}=-\frac{z^3+c}{3z^2}.
\end{equation}

Roots of the polynomial equation~(\ref{f3}) are the stable nodes of the 
ODE~(\ref{ode}). At the same time these roots are the attractors of the 
Newton and Euler iteration process~(\ref{newt}), at least, when the discretization 
step~$h$ is small enough. 
The boundary between basins of attraction of the iteration process~(\ref{newt})~--- the fractal Newtonian Julia 
set~--- corresponds to a separatrix of~(\ref{ode}), which should be smooth 
when $f(z)$ is defined by~(\ref{fun}) (see 
Fig.~\ref{fig0}). Fractalization of the separatrix occures due to
the Euler discretization. This is a numerical artifact, which can be considered as neglectable for practical applications at $h\rightarrow 0$ .

While in general discretization of flow dynamical systems caused by numerical time-integration can lead to emergence of solutions which do not represent the dynamics
of the original system and manifest themselves in changes of the phase space structure, as in example discussed above, or changes in bifurcation diagrams etc. (see, e.g.,~\cite{Pa1,Pa2,Pa3} and references there), this problem can be considered from another angle: such discretization of well-known flow systems can be regarded as a fruitful approach, which is widely used in modern nonlinear theory and allows to generate new model maps (see, e.g.,~\cite{ar1993,za2007,Kuz2009,Ad2013} and references there). Due to emergence of numerical artifacts mentioned above dynamical systems generated this way can demonstrate various nontrivial phenomena. The discretization step~$h$ is usually defined in such models in a wide range~--- moreover, it can be complex~\cite{Peitgen}.
This approach can also establich a background for introducing and considering of a new class of systems, one example from which is proposed in present work.

Simple construction~(\ref{newt}), being considered as an abstract 
dynamical system, allows a wide spectrum of generalizations and 
parametrizations. Types of dynamical behavior and obtained phenomena can 
also be rather diverse. Particularly, the implicit dynamics, when every 
point in the phase space of the system has both several images and several 
preimages~\cite{Osb,Bu2,Me1}, is also possible. Such implicit correspondences have wide spectrum of applications 
besides implicit numerical schemes of equations solving. Implicit 
functions can occur in problems of reconstruction of a multidimensional object (or system) 
from its projection~\cite{Di1}, in the theory of generalized 
synchronization~\cite{Pi1}, in economics~\cite{ke2008,Ga1}, computer graphics~\cite{Sc3},
chaos control techniques~\cite{Hi1}, topology~\cite{Vl1}.

In the present paper we try 
to give an example of generalization of the iteration process~(\ref{newt}) and to investigate obtained system 
from the point of view of theoretical nonlinear dynamics.
In the section 1 we present the procedure of deriving an implicit map using the modified Euler method. In the section 2 we analyze the fractalization 
of both unstable and stable invariant sets of such exotic system.

\section{Basic model}

Among the numerical recipes of the ODE integration the
semi-implicit Euler method is listed:
\begin{equation}\label{seul}
z_{n+1}=z_n+h(f(z_n)+f(z_{n+1}))/2.
\end{equation}
Let us generalize this scheme by parameterization:
\begin{equation}\label{seu1l}
z_{n+1}=z_n+h((1-\alpha) f(z_n)+\alpha f(z_{n+1}).
\end{equation}
In case when $f(z)$ chosen in form~(\ref{fun}) this equation can be rewrited 
as following:
\begin{equation}\label{impl}
(\alpha h+3)z_{n+1}^3 z_n^2+((1-\alpha) h-3)z_{n+1}^2 z_n^3
+(1-\alpha)h cz_{n+1}^2+\alpha hcz_n^2=0.
\end{equation}
This is also some iteration process, but, in contrast 
to~(\ref{eul}), not traditional for the nonlinear dynamical system theory.
This is an implicit map with the evolution operator looking like
\begin{equation}\label{psi}
\Psi(z_{n+1},z_n)=0.
\end{equation}
Both forward and backward iterations of this map are defined by multi-valued functions, 
\begin{equation}\label{forw}
z_{n+1}=\Psi_{+}(z_n)
\end{equation}
and 
\begin{equation}\label{back}
z_n=\Psi_{-}(z_{n+1})
\end{equation}
respectively.

Two examples of implicit maps are described in~\cite{Osb,Isa}.
Below we are trying to study the implicit dynamics on the new example of such map~(\ref{impl}).

\section{Numerical simulation}
\subsection{Repellers}
\begin{figure}
\includegraphics[width=\textwidth]{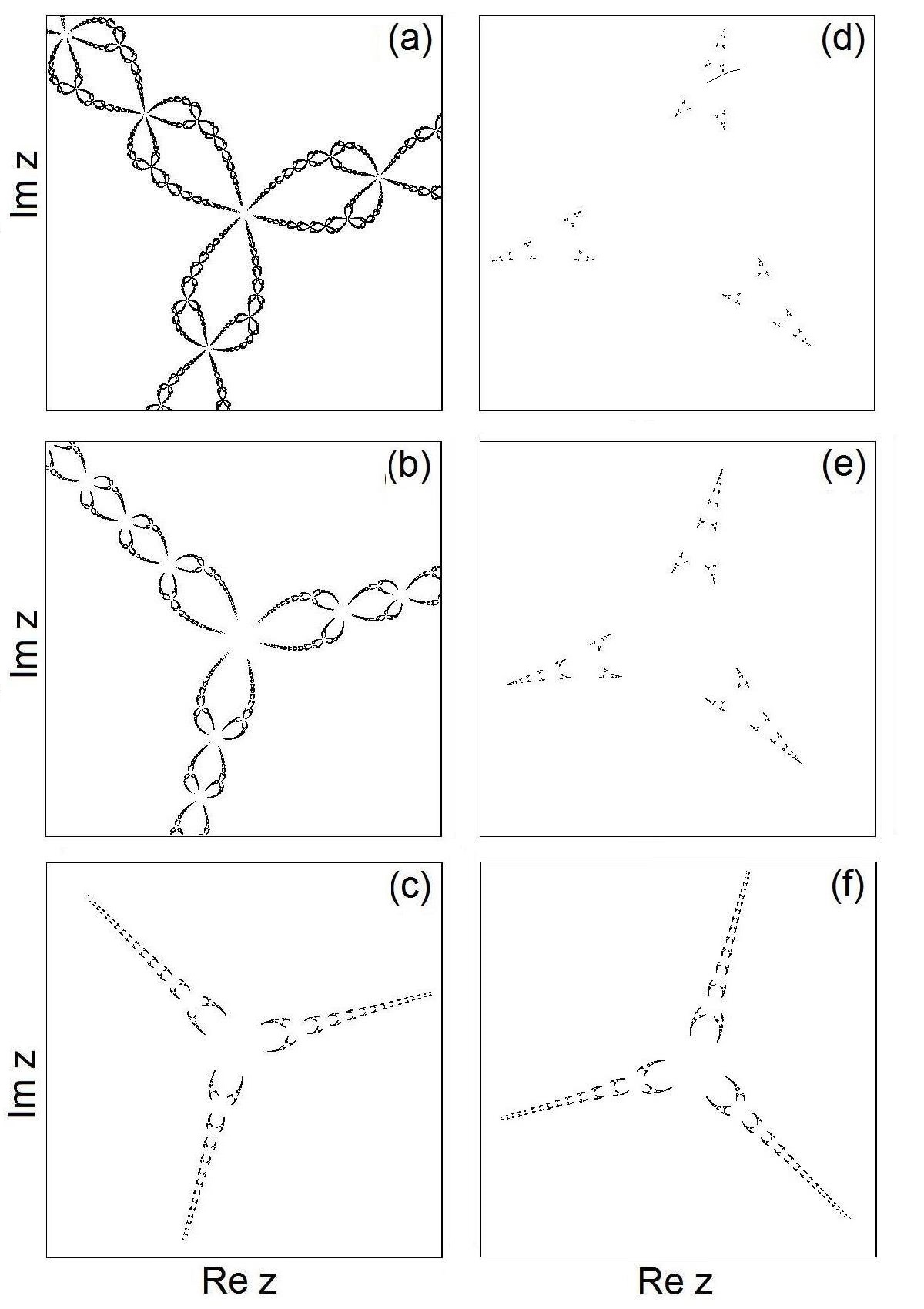}
\caption{The repellers in the phase space of the map~(\ref{impl}) 
with $h=1.0$~(a), $h=0.5$~(b), $h\rightarrow +0$~(c) and 
$h=-1.0$~(d), $h=-0.5$~(e), $h\rightarrow -0$~(f). Parameter 
$\alpha$ is equal to zero. Parameter $c$ is not essential, 
here and further it is fixed being equal to $(1-i)/2$. } \label{rep1}
\end{figure}

We will start investigation from studying backward dynamics of the map~(\ref{impl}) at $\alpha=0$. In this case the map is single-valued time-forward and multi-valued time-backward.
We will study structure of its repellers, which form boundaries of basins of attraction. It
is worth mentioning here that since the map~(\ref{impl}) is defined by the cubic
polynomial, solutions of the equation~(\ref{back}) can be found analytically.    
The repellers of the map~(\ref{impl}) at different values of  $|h|\le1$
are shown in the Fig.~\ref{rep1}.
At $h=1$ we obtain the classical Newtonian Julia set, other positive values of $h$
correspond to transformations of this fractal (see Fig.~\ref{rep1}a,b). Repellers in this case still define
boundaries of areas of convergence of the Newton method~(\ref{newt}) to different roots of~(\ref{f3})~--- or of 
the Euler method to different nodes of the ODE~(\ref{ode}). At negative values of $h$ the process of search of repellers of the map~(\ref{impl}) corresponds to solving the ODE~(\ref{ode}) with the time reversed with the Euler method. The result of this process is a fractal set similar to the Sierpi\'{n}sky gasket (see Fig.~\ref{rep1}d,e). $h=0$ is a degenerate case corresponding transition between these two situations (see Fig.~\ref{rep1}c,f).

\begin{figure}
\includegraphics[width=\textwidth]{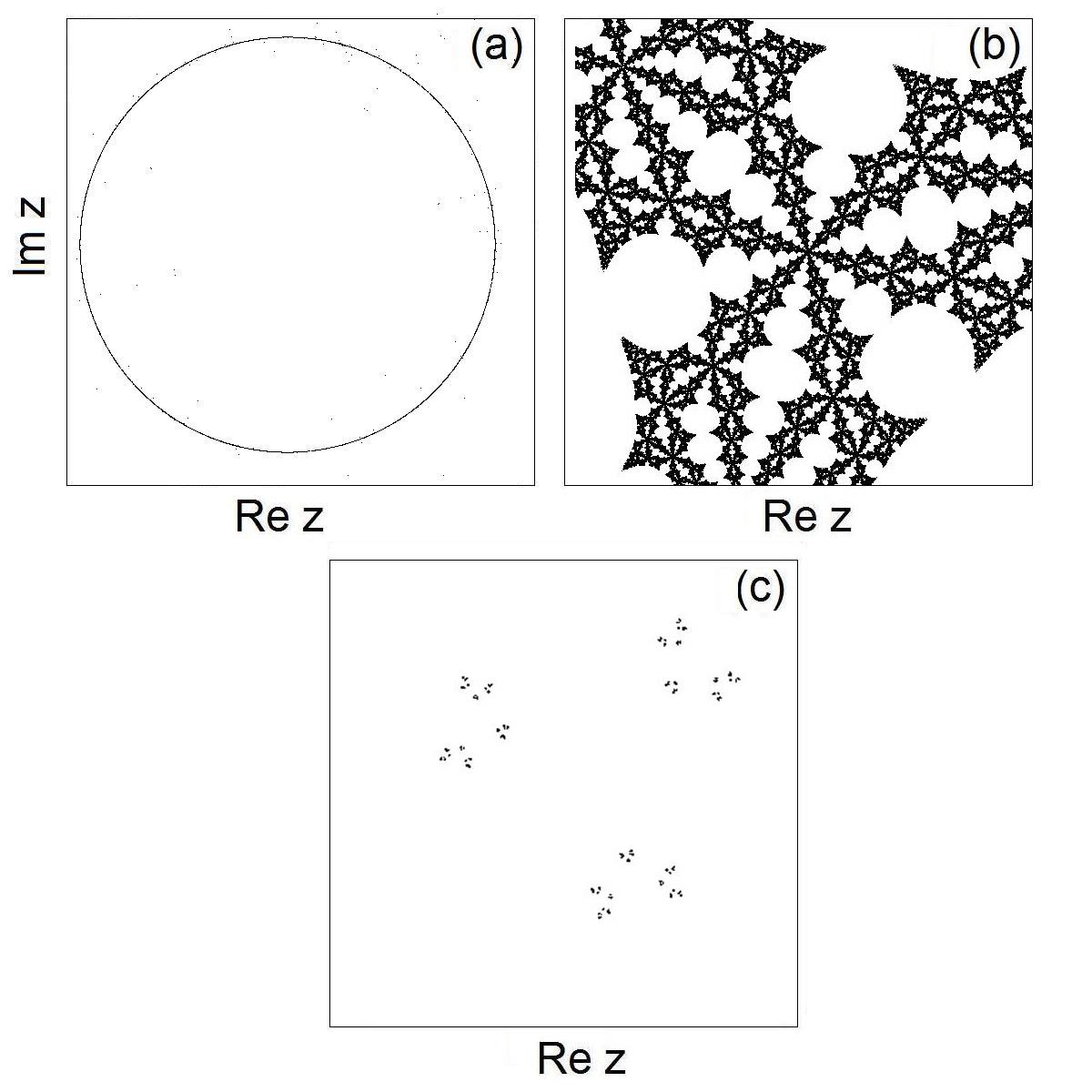}
\caption{The repellers in the phase space of the map~(\ref{impl}) 
with $\alpha=0$ and different values of $h$: $3$~(a), $5$~(b), $7$~(c).} \label{rep2}
\end{figure}

The repellers for $h>1$ are shown in the
Fig.~\ref{rep2}, here $h=3$ (Fig.~\ref{rep2}a) is also a degenerate case~--- which follows from 
the structure of function~(\ref{impl}), where one of the terms becomes in this case equal to 0. 

\begin{figure}
\includegraphics[width=1.0\textwidth]{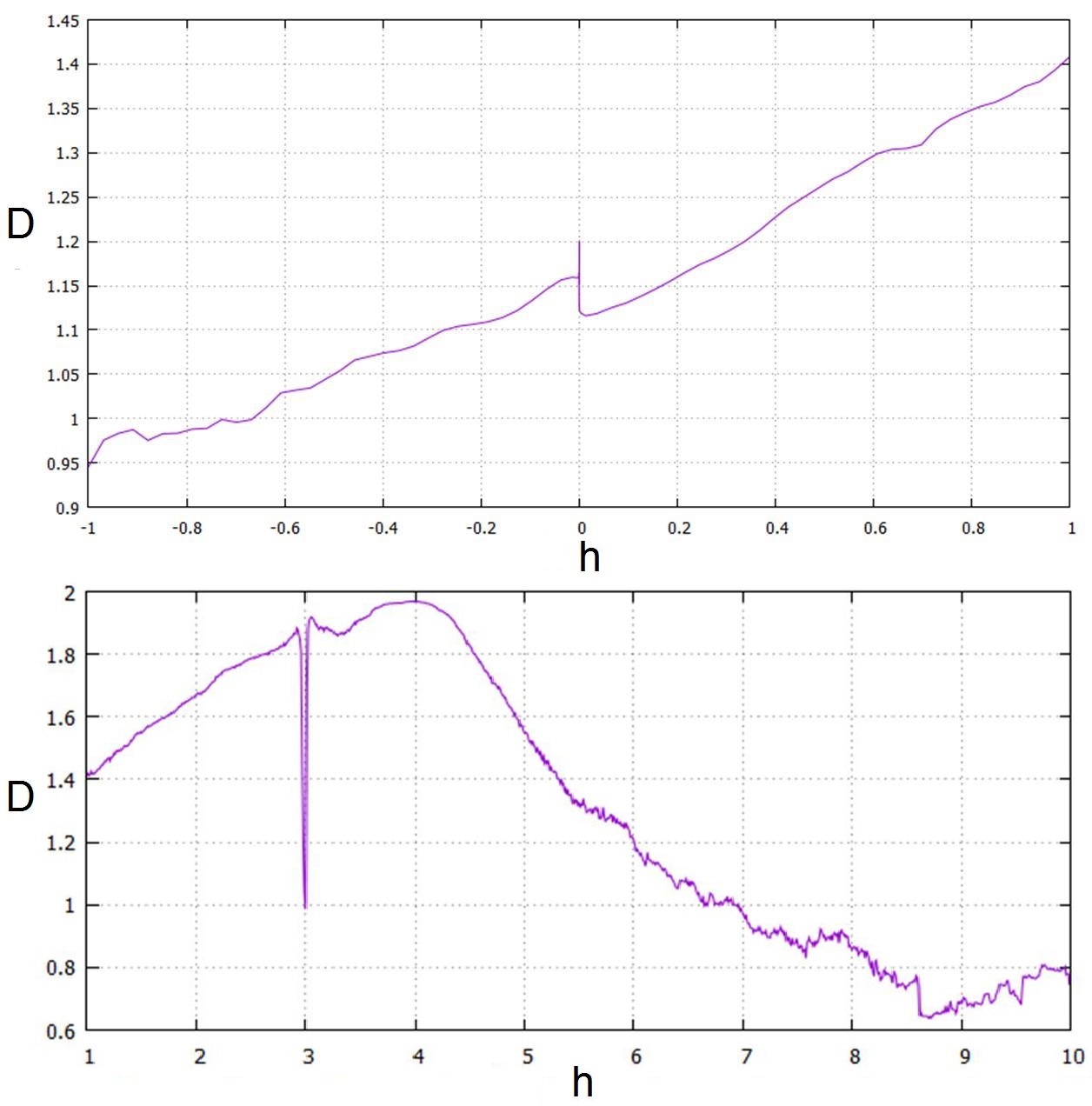}
\caption{Box-counting dimension of the repellers of the 
implicit map~(\ref{impl}) with~$\alpha=0.0$ for $|h|\le 1$ (upper panel) and $1 \le h \le 10$ (bottom panel).} \label{fig-frac}
\end{figure}

A useful tool for quantitative analysis of the phase space structure
transformations is a fractal dimension of the basin boundaries.
Fig.~\ref{fig-frac} demonstrates the dependence of the box-counting dimension on the parameter $h$ value.
In the vicinity of $h=0$ an abrupt change of the value of dimension occurs, which indicates a phase transition of the Julia set.
In the vicinity of the degenerate case $h=3$ the fractal dimension value tends to 1~--- 
it corresponds to degeneration of the Julia set, which looks in this case like a
smooth circle (Fig.~\ref{rep2}a). In the region near $h=4$ the value of dimension 
grows almost up to $2$, and Julia set almost becomes a fat fractal (Fig.~\ref{rep2}b).
When $h\geq 7$, the value of dimension decreases down to values below 1.
The attracting invariant set undergoes here crisis, and basin boundaries degenerate to the fractal dust.

\subsection{Attractors}

In general case every point in the phase space of the implicit map~(\ref{impl}) has three images~--- roots of qubic 
polynomial equation. Time-forward dynamics of such map is, as backward dynamics also, not single-valued.
To study time-forward dynamics it seems productive to apply methods which are usually employed for an analysis of repellers.
Particularly, we apply the ``chaos game'' algorithm~\cite{Peitgen} in order to find attracting chaotic trajectories. To find periodic trajectories
we choose the roots of~(\ref{impl}) at each iteration according to a periodic sequence. We construct symbolic codes, 
consisting of characters <<1>>, <<2>> and <<3>>~--- which correspond to the choice of the first, the second or 
the third root respectively. The period of dynamical orbit should be in this case
equal or larger than a minimal period of such a sequence.

\begin{figure}
\includegraphics[width=\textwidth]{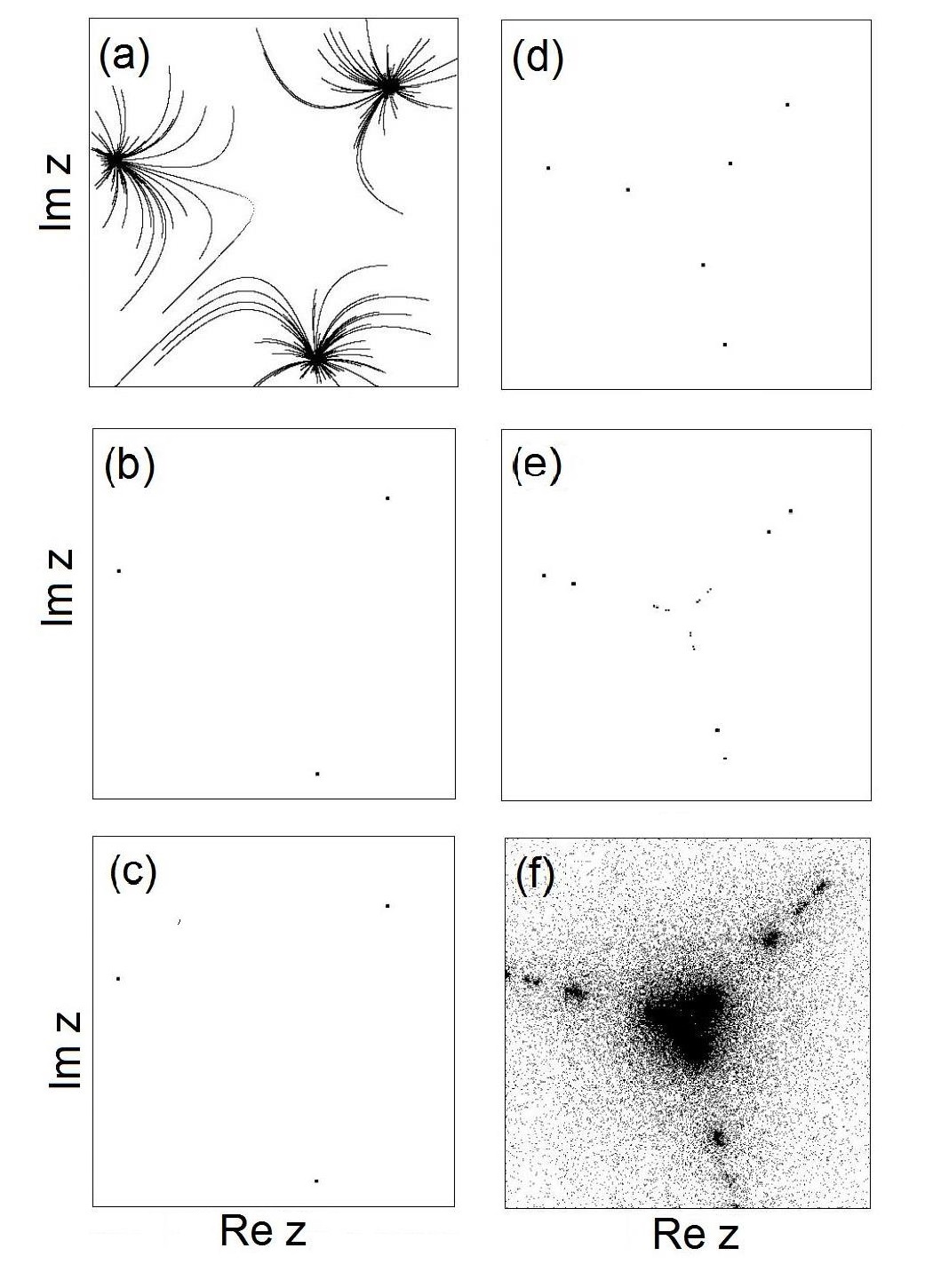}
\caption{Transformation of the attracting invariant sets of the
implicit map~(\ref{impl}) with~$\alpha=0.0$ and with $h\rightarrow 0$~(a), 
$h=0.5$~(b), 1.0~(c), 2.15~(d), 2.75~(e), 2.792~(f).} \label{fig-att}
\end{figure}

Let us start from the illustration of the evolution of attractors 
in special case $\alpha=0$, which is shown in the Fig.~\ref{fig-att}. This situation corresponds to the use of the explicit Euler method, and the 
forward-time dynamics is in this case single-valued. Fig.~\ref{fig-att}a represents three attracting nodes at $h\rightarrow 0$,  
while an increase of parameter $h$ causes several period doubling bifurcations (Fig.~\ref{fig-att}c-e), and the transition to chaos occurs (Fig.~\ref{fig-att}f).

\begin{figure}
\includegraphics[width=\textwidth]{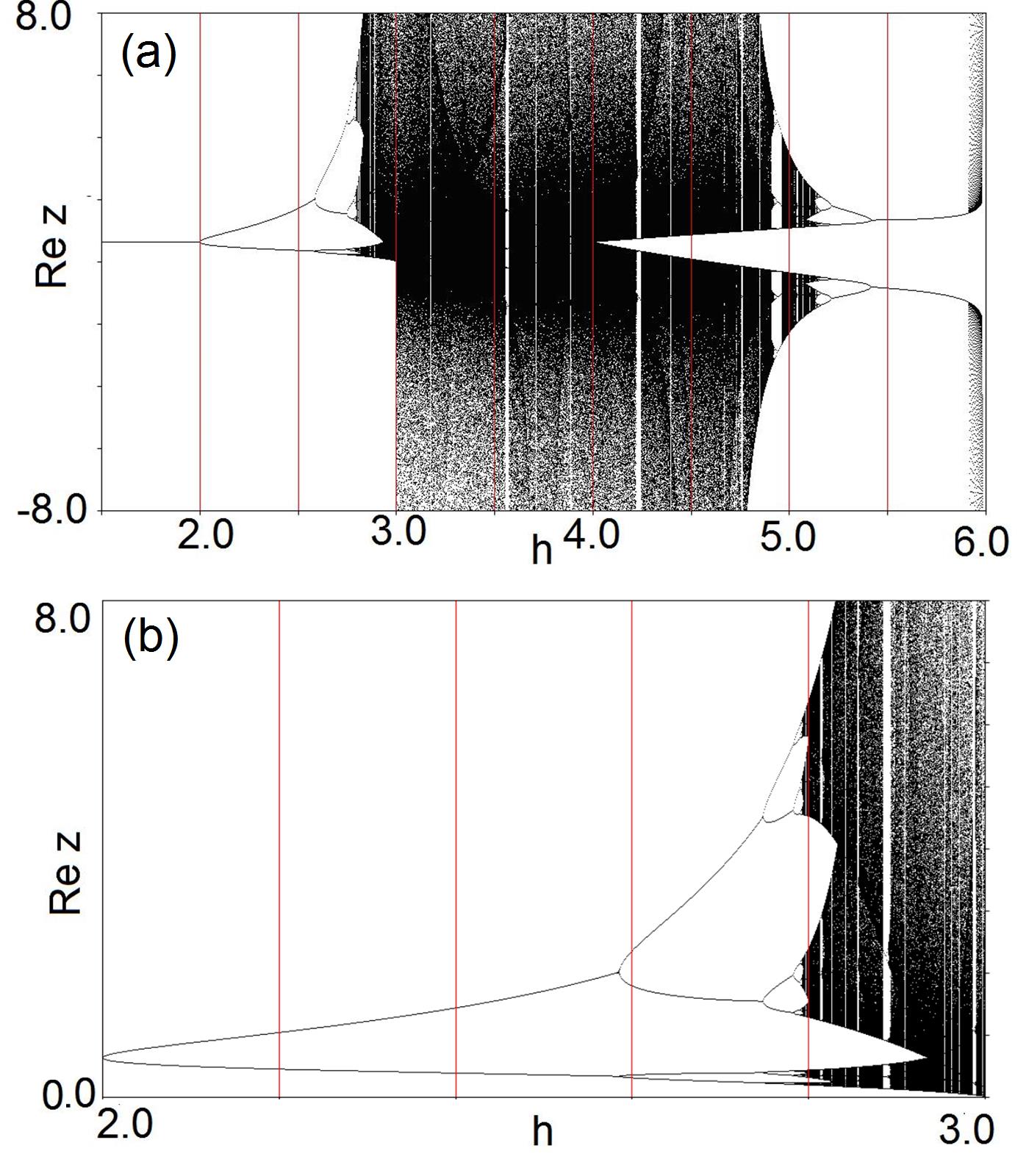}
\caption{The bifurcation diagram for one of attracting 
invariant sets of the implicit map~(\ref{impl}) with~$\alpha=0.0$~(a) 
and its enlarged fragment~(b).} 
\label{fig-tree}
\end{figure}

The bifurcation diagram shown in the Fig.~\ref{fig-tree} gives more complete picture of the forward-time behavior of the 
system~(\ref{impl}). Here the real part of the complex variable $z$ on the attracting 
invariant set is plotted versus parameter $h$. At this picture the evolution of one of three attracting fixed points is demonstrated.
It undergoes several period-doubling bifurcations, transition to chaos and back to the periodic regime, and finally is destroyed via crisis. In the vicinity of the point $h=4$, where, as we mentioned above, Julia set almost becomes a fat fractal, time-forward dynamics is chaotic.

Next Figure~\ref{fig-char} demonstrates the picture of dynamical 
regimes on the parameter plane $(h, \alpha)$. When $\alpha\neq 0$, the map~(\ref{impl}) becomes an implicit one, which means that its dynamics is now muti-valued both time-forward and time-backward.
\begin{figure}
\includegraphics[width=\textwidth]{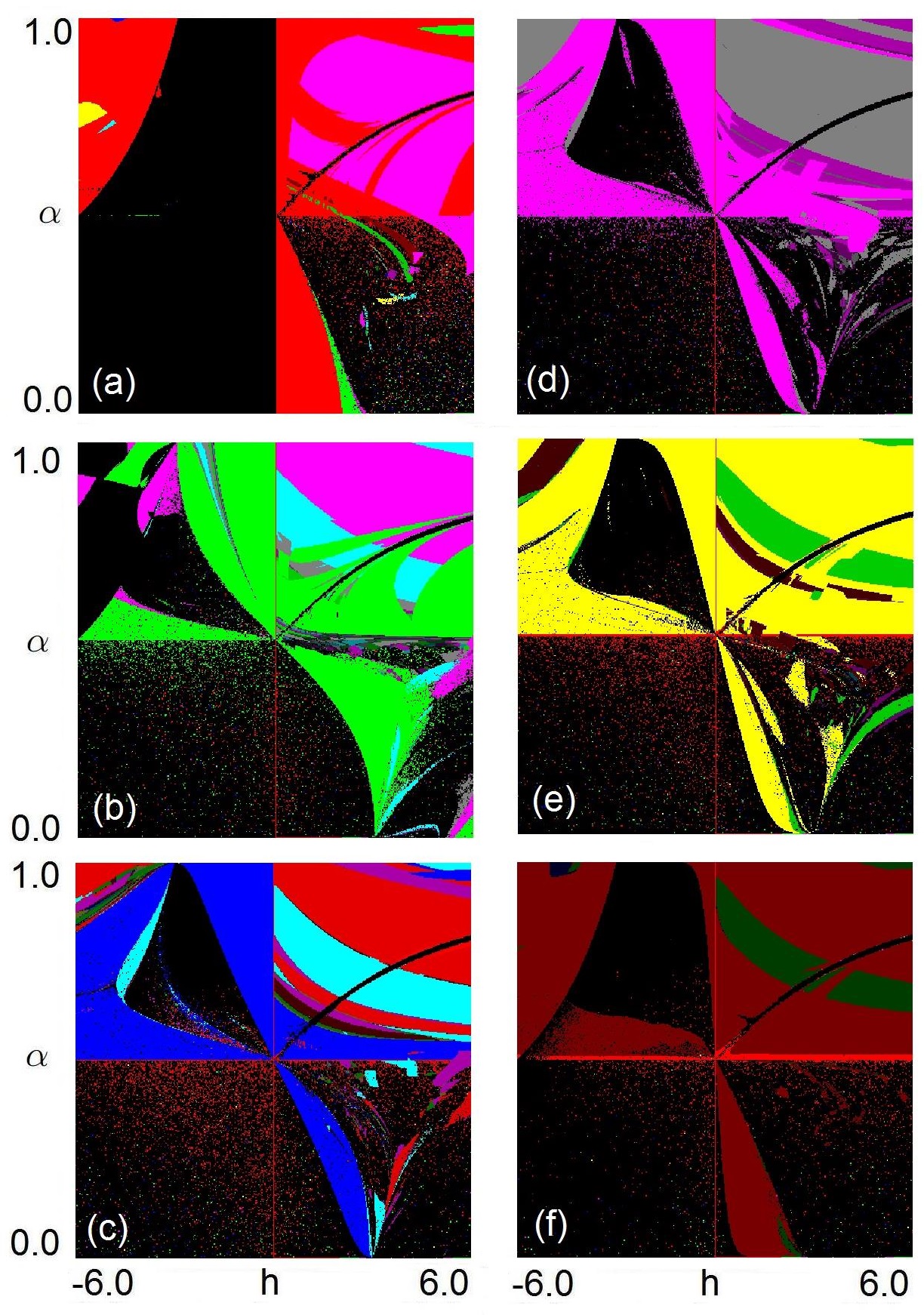}
\caption{The charts of dynamical regimes of the 
implicit map~(\ref{impl}) for the following periodic sequences 
of time-forward roots choice: $1$~(a), $12$~(b), $112$~(c), $1112~$(d), 
$11112$~(e), $1(\times 8)2$~(f).} \label{fig-char}
\end{figure}
  Except the simplest case, when the symbolic sequence has period 1 (Fig.~\ref{fig-char}a), charts for different periodic sequences demonstrate similar features. Areas of periodic dynamics in the quadrant $(h>0, \alpha<0.5)$ have typical ``tongue'' shape with spikes in the point $\alpha=0, h=3$, where the map~(\ref{impl}) is degenerate. Another typical feature is partial symmetry of the parameter plane: borders of areas of aperiodic dynamics are in many cases symmetrical with respect to the point $(h=0, \alpha=0.5)$ in quadrants $(h>0, \alpha<0.5)$ and $(h<0, \alpha>0.5)$, and points with periodic dynamics are symmetrical to points with aperiodic dynamic, especially for $|h|\le 3$. It is a consequence of specific symmetry of the map~(\ref{impl}), which is invariant with respect to the transformation $h\rightarrow -h, \alpha \rightarrow 1 - \alpha, z_n\leftrightarrow z_{n+1}$.

\section{Conclusion}

In this paper we present a short preliminary view on the dynamics 
of one example of implicit systems, namely, the map~(\ref{impl}). 
We demonstrate some approaches for studying of such systems.
Advanced investigation of this implicit map should clarify the structure of 
its invariant sets. It seems a promising and very interesting direction of research, since in implicit systems, in contrast to traditionally studied explicit ones, an infinite number of trajectories can coexist in forward time, which makes their attracting invariant sets very complicated. Moreover, complexification of the parameter $h$  in the map~(\ref{impl}) leads to the possibility of a situation, when $\Psi(z_{n+1},z_n)=-\Psi^*(z_n, z_{n+1})$. \footnote{Here $\Psi^*(z_{n+1},z_n)=(\Psi(z^*_{n+1},z^*_n))^*$.} For the map~(\ref{impl}) this happens at $h=\pm i$, $\alpha=1/2$. This situation, defined in~\cite{Osb,Isa} as generalized unitarity, manifests the emergence of phenomena typical for Hamiltonian and almost Hamiltonian systems. In this context the implicit maps, being an artificial construct, can help to describe strong multistability, mixed dynamics and other complex phenomena of nonlinear dynamics.

%
%
\bibliographystyle{splncs04}
\bibliography{mybibliography}

\end{document}